\documentclass[%
twocolumn,
 amsmath,amssymb,
showpacs
]{revtex4-1}

\usepackage[T1]{fontenc}
\usepackage[latin9]{inputenc}
\usepackage{geometry}

\usepackage{float}
\usepackage{amsmath}
\usepackage{amssymb}
\usepackage{graphicx}
\usepackage{epstopdf}



\begin{document}

\title{Calculation of molecular free energies in classical potentials}
\author{Asaf Farhi}
\email{asaffarhi@post.tau.ac.il}
%\author{David J. Bergman}
%\email{bergman@post.tau.ac.il}
\affiliation{
Raymond and Beverly Sackler School of Physics and Astronomy,
Faculty of Exact Sciences,
Tel Aviv University, IL-6997801 Tel Aviv, Israel
}
\affiliation{Physics Department, Weizmann Institute of Science, Rehovot 76100, Israel}
\date{\today}
\begin{abstract}
Free energies of molecules can be calculated by quantum computations or by normal mode classical calculations. However, the first can be computationally impractical for large molecules and the second is based on the assumption of harmonic dynamics. We present a novel, accurate and complete calculation of molecular free energies in standard classical potentials. In this method we transform the molecule by relaxing potential terms which depend on the coordinates of a group of atoms in that molecule and calculate the free energy difference associated with the transformation. Then, since the transformed molecule can be treated as non interacting systems, the free energy associated with these atoms is analytically or numerically calculated. This two-step calculation can be applied to calculate free energies of molecules or free energy difference between (possibly large) molecules in a general environment. We suggest the potential application of free energy calculation of chemical reactions in classical molecular simulations.

\end{abstract}

\pacs{05.10.-a,02.70.Ns,83.10.Rs}

\maketitle

\section*{Introduction}
Free energy calculations have a variety of applications which include binding, solvation and chemical reactions. For the applications of solvation and binding equilibrium and non-equilibrium methods are used.
In equilibrium methods (alchemical transformations) a molecule is transformed into another through non realistic intermediate systems. The average of the derivative of the interpolating Hamiltonian is calculated for each intermediate system 
and by numerically integrating, the free energy associated with the transformation is calculated. The free energies of these transformations are then used to calculate the free energy of the desired molecular process \cite{zuckerman2011equilibrium,chipot2007free,frenkel1996understanding,binder2010monte,allen1987computer}. In non equilibrium methods the work in the process of switching between the two Hamiltonians is measured \cite{jarzynski1997nonequilibrium,crooks1999path}. For the application of chemical reactions the free energy difference between the reactants and products needs to be calculated. Free energies of molecules can be calculated by quantum computations \cite{jorgensen1987priori} or by normal mode classical calculations \cite{lii1989molecular} (see Entropies Section). While QM calculations are highly accurate they are computationally very demanding for large molecules. Normal mode calculations can be very efficient but are based on harmonic dynamics \cite{hayward1995collective}.

%These methods include Jarzynski relation \cite{jarzynski1997nonequilibrium} and its subsequent generalization by Crooks \cite{crooks1999path}.

Molecular modeling includes covalent bond, bond angle, dihedral angle, improper dihedral angle, electrostatic and vdW potentials (see \cite{mayo1990dreiding,phillips2005scalable,hess2008gromacs} and Supplementary Material).
Covalent bond, bond angle and dihedral angle terms depend on the coordinates of two, three and four nearest covalently linked atoms respectively. Electrostatic and vdW potentials are between every atom pair in the system.
%The covalent bond, bond angle and dihedral angle terms can be expressed in terms of the spherical variables $r$, $\theta$ and $\phi$ defined with respect to the relevant atoms.   

In previous studies, the free energy associated with the covalent bond and bond angle terms in molecules and dihedral angle term in the context of restraints were \emph{directly} calculated by assuming that they are harmonic (quadratic) and by using the rigid rotor \emph{ approximation} and the HJR technique \cite{boresch1999role,boresch2003absolute}. However, the free energies associated with dihedral potential terms in the realistic non-harmonic potentials and with non-harmonic covalent bond and bond angle terms \cite{malde2011automated} have not been calculated. In addition, it has not been suggested to calculate molecular free energies of submolecules with coupled bond angle degrees of freedom such as the methyl group.

Here we present a \emph{direct} free energy calculation which is \emph{exact} and does not require the potential terms to be harmonic (see also Ref. \cite{ToBePublished}). We show that the partition function associated with a dihedral angle in molecules can be decoupled from the partition function of the rest of the system independently of the potential function (even though the dihedral potential term depends on coordinates of atoms in the system) and calculate analytically free energies of realistic dihedral terms. Moreover, we show that free energy associated with submolecules with coupled bond angle degrees of freedom such as the methyl group can be calculated accurately. 

In Section I we present the first step of the calculation in which we transform the molecule by relaxing some potential terms of a group of atoms in that molecule and calculate the free energy difference associated with the transformation. In Section II we present the second step of the calculation in which the free energy associated with the remaining potentials  (which depend on the relative spherical coordinates) of this group of atoms is calculated analytically or numerically. In Section III we suggest the potential application of free energy calculation of chemical reactions. The demonstartions of the method are explained in the Demonstration section.

Alchemical free energy transformation (as in Section I) is a standard procedure in molecular free energy calculations \cite{hess2008gromacs,phillips2005scalable,cuendet2012alchemical,salomon2013overview,mugnai2012thermodynamic} and is usually used in the context of solvation and binding or in the context of chemical reactions in combination with QM calculations \cite{jorgensen1987priori}. Here we apply it in the context of free energy calculations of molecules or submolecules in a general environment, where the second calculation is used to obtain the free energy difference between two molecules with submolecule in common. We combine the calculation with novel direct free energy calculations explained in Section II, followed by the novel application of free energy of chemical reactions presented in Section III. Here we \emph{do not} need to assume harmonic dynamics.

\section{The transformation}
We now explain the first step of the calculation in which we we transform the molecule by relaxing potential terms that depend on the coordinates of a group of atoms in that molecule and calculate the free energy difference associated with the transformation using Thermodynamic Integration (TI) \cite{frenkel1996understanding,kirkwood1935statistical,straatsma1991multiconfiguration}.  Alternatively, it can be performed using methods such as Exponential Averaging/ Free Energy Perturbation (FEP) \cite{zwanzig1954high}, Bennett Acceptance Ratio (BAR)
\cite{bennett1976efficient} and Weighted Histogram Analysis Method
\cite{kumar1992weighted} (WHAM). We denote the Hamiltonian with the terms that are removed in the transformation by $H_r$ and the Hamiltonian with the other terms by $H_c$. We define the Hamiltonian $H_r$ as the improper dihedral, electrostatic and vdW terms that depend on the coordinates of the group of selected atoms.
The $\lambda$ dependent Hamiltonian that transforms between the systems can be written as follows:
\begin{equation}
H\left(\lambda\right)=\lambda
H_r+H_c,\nonumber
\end{equation}
where $\lambda \in [0,1]$.
\begin{figure}[h]
 \centering
\includegraphics[width=8cm]{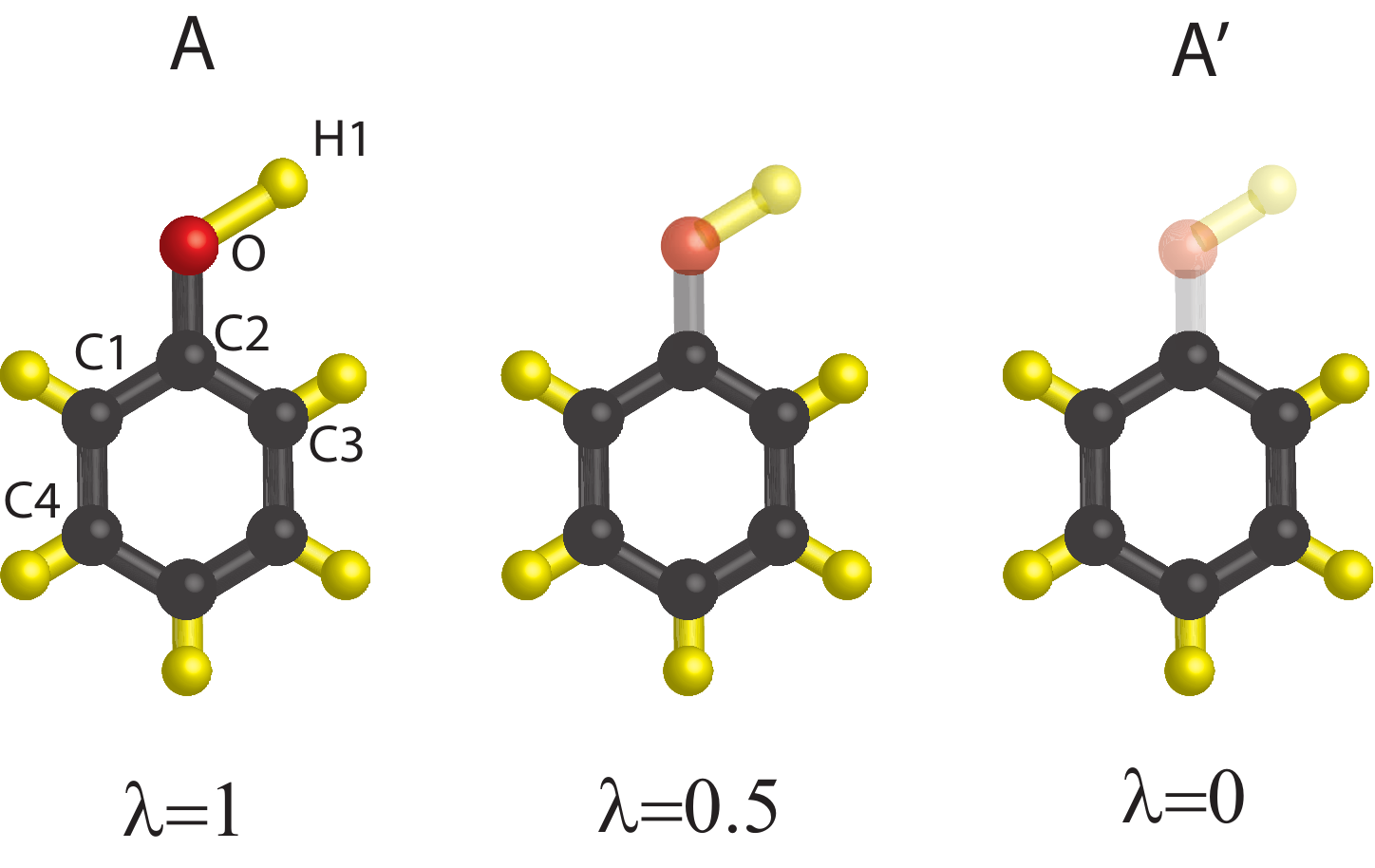}
 \caption{Systems in the transformation of phenol at $\lambda=0,0.5,1$. The semi transparent/ transparent atoms represent atoms with interactions partly/ totally removed}
 \label{fig:phenol}
\end{figure}
The free energy difference between the original and transformed systems can be written as:
$$\Delta F_{A\rightarrow A'}=-k_{B}T\int_{0}^{1}\frac{d \ln Z_{A}\left(\beta,\lambda\right)}{d\lambda}d\lambda=\int_{0}^{1}\left\langle H_{r}\right\rangle d\lambda,$$
%\begin{eqnarray}
%\Delta F_{A\rightarrow A'}&=&-k_{B}T\left[ \ln Z_{A}\left(\beta,\lambda=1\right)-\ln Z_{A}\left(\beta,\lambda=0\right)\right]\nonumber\\
%&=&-k_{B}T\int_{0}^{1}\frac{d \ln Z_{A}\left(\beta,\lambda\right)}{d\lambda}d\lambda=\int_{0}^{1}\left\langle H_{r}\right\rangle d\lambda, \nonumber
%\end{eqnarray}
where $A'$ is the transformed system in which $\lambda=0$.
Simulating intermediate systems at $\lambda$ values in the range $[0,1]$ and calculating $\left\langle H_{r}\right\rangle$, followed by numerical integration, will enable us to calculate the free energy difference.
It is noted that this calculation is valid for any environment (gas, solvent etc.) since the interactions of the group of atoms with the environment are also relaxed in the transformation. In addition, soft core potentials are used in the Vdw and electrostatic interactions to avoid divergences of 
$\left\langle H_{r}\right\rangle$ at $\lambda\rightarrow0$ \cite{beutler1994avoiding}.

As an example we present in Fig. \ref{fig:phenol} three systems in the transformation of the molecule phenol with the selected atoms O and H1.
The Hamiltonian $H_r$ includes the vdW and electrostatic interactions of the atoms O and H1 with all the atoms in the system, and the improper dihedral term $\phi_{C1-C2-O-C3}$.

\section{The remaining free energy difference}
We now turn to explain how the free energy associated with the selected atoms can be calculated.  
We first switch to relative coordinates and then to spherical coordinates. 
\begin{eqnarray}
d\Omega&=&\prod_{i=1}^{n}d\mathbf{r'}_{i}=d\mathbf{r'}_{1}\prod_{i=2}^{k}d\mathbf{r{}}_{i}\prod_{i=k+1}^{n}d\mathbf{r}_{i}\nonumber\\
&=&d\mathbf{r'}_{1}\prod_{i=2}^{k}d\mathbf{r}_{i}\prod_{i=k+1}^{n}r_{i}^{2}\sin\theta_{i}dr_{i}d\theta_{i}d\phi_{i},\nonumber
\end{eqnarray}
where $\mathbf{r}_{i}\equiv \mathbf{r}'_{i}-\mathbf{r}'_{i-1}$,
which will be chosen as the position of atoms relative to a covalently bound atom. $k+1$ represents the first atom in the group of selected atoms (e.g the atom O in the example of phenol). 
%In the case that the molecule is unmatched $k+1$ will be the first atom.

The transformed molecule A' is first divided into elements of standard covalent bonds, bond junctions, dihedral terms 
and of more complex structures that include molecular rings. The dihedral potential term depends on the angle between two planes which is equal to the $\phi$ angle in spherical coordinates. This correspondence can be understood by recalling the definition of the angle between two planes as the angle between the vectors in these planes that are perpendicular to the intersection line of these planes.
Hence, the covalent bond, bond angle and dihedral angle terms can be expressed in terms of the spherical variables $r$, $\theta$ and $\phi$ defined with respect to the relevant atoms.

For the example of phenol we can write the partition function of transformed phenol as follows
\begin{gather}
 Z_{A'}=\int e^{-\beta H\left(\mathbf{ \Omega} \right)}e^{-\beta H_{r_{O}}(r_{O})}  e^{-\beta H_{r_{H}}(r_{H})} \times  \\
 e^{-\beta H_{\theta_{H}}(\theta_{H})}\sin\theta_{H}e^{-\beta H_{\phi_{H}}\left(\phi_{H}\right)}d\mathbf{ \Omega} r_{O}^2 dr_{O}r_{H}^2 dr_{H}d\theta_{H} d\phi_{H}
  \nonumber,
\label{partition_transformed_phenol}
\end{gather}
where the degrees of freedom of atom H and O are denoted by $r_{\mathrm{H}},\theta_{\mathrm{H}},\phi_{\mathrm{H}}$ and   $r_{\mathrm{O}},\theta_{\mathrm{O}},\phi_{\mathrm{O}}$ respectively. All the degrees of freedom of all the other atoms in the system  (molecule+solvent molecules in the case of explicit solvent) \emph{and} $\theta_{\mathrm{O}},\phi_{\mathrm{O}}$  are denoted by $\mathbf{\Omega}$.  We define $\theta_{H}$ and $\phi_{H}$ as follows: $\theta_{H}\equiv\theta_{H1-O-C2},\phi_{H}\equiv\phi_{H1-O-C2-C1}$. Our goal in this example is to calculate the free energy associated with all the degrees of freedom except for the ones denoted by $\mathbf{\Omega}$. The free energy of the subsystem with the degrees of freedom denoted by $\mathbf{\Omega}$ can for example then cancel out with an identical subsystem when calculating the free energy difference between two molecules with a submolecule in common.  Note that for example $\phi_{H}$ \emph{does} depend on the coordinates of the $C1$ atom which is included in $\mathbf{\Omega}$ and therfore the decomposition of the partition functions is not trivial.
 
As will become clear, the integration in each element is independent of the others. Thus the
integrals can be performed separately and then multiplied to yield
the partition function and hence the free energy difference.

We write the free energy factors explicitly:
\subsection*{Covalent bond}
The partition function of a covalent bond represented by a harmonic (quadratic) potential can be written as follows (for non-quadratic terms integrated analytically see Ref. \cite{ToBePublished}):
\begin{widetext} 
\begin{equation}
\label{eq:covalent_bond}
Z_c=\frac{1}{l^3} \int e^{-\frac{\beta k_c\left(r-d\right)^{2}}{2}}r^{2}dr=\frac{\pi^2\left[\left(2d^{2}\beta k_c+1\right)\left(\mathrm{erf}\left(d\sqrt{\beta k_c}\right)+1\right)+2\sqrt{\pi}e^{-\beta k_cd^{2}}d\sqrt{\beta k_c}\right]}{l^{3}\left(\beta k_c\right)^{3/2}},
\end{equation}
\end{widetext}
where $l$ is an arbitrary length ($l^3$ cancels out in comparisons). To account for the bond-dissociation energy/bond energy one has to differentiate between the potential at $r=0$ and $k_{c}\rightarrow0$ (free atoms) either by introducing a factor \cite{zavitsas1987quantitative,blanksby2003bond,allinger1989molecular} or by using potentials such as the Morse potential (numerical integration) \cite{morse1929diatomic}.% Analytic or numeric integrations with these potentials can be readily performed.
\subsection*{Two Bonds Junctions}
When considering the case of a Linear/Bent molecular shapes, it can
be seen that when varying the bond angle, the rest of
the molecule moves as a rigid body. Since the spherical coordinates
representation includes three independent variables, varying a bond 
angle is decoupled from all the other degrees of freedom of the molecule.
Hence the calculation of free energy factor associated with the bond angle 
potential term is straightforward. For a harmonic (quadratic) term we can write :
%\begin{equation}
%V_{b}\left(\theta\right)=\frac{1}{2}k_{\theta}\left(\theta-\theta_{0}\right)^{2},  \nonumber\end{equation}
%The integration over the corresponding degree of freedom can be written as:
\begin{widetext}
\begin{gather}
Z_{b}=\int e^{-\beta V_{b}}\sin\theta d\theta=\int e^{-\frac{\beta k_{\theta}}{2}\left(\theta-\theta_{0}\right)^{2}}\sin\theta d\theta=\frac{1}{2\sqrt{\beta k_{\theta}}}e^{-i\theta_{0}-\frac{1}{2\beta k_{\theta}}}\times \nonumber \\
\label{eq:bond_angle}
 \sqrt{\frac{\pi }{2}} \left(i \mathrm{erf}\left[\frac{i-\theta_{0}\beta k_{\theta}+\beta k_{\theta} \pi }{\sqrt{2\beta k_{\theta}}}\right]+\mathrm{erfi}\left[\frac{1+i \theta_{0}\beta k_{\theta}}{ \sqrt{2\beta k_{\theta}}}\right]-i e^{2 i \theta_{0}} \left(\mathrm{erf}\left[\frac{i+\theta_{0}  k_{\theta}}{\sqrt{2\beta k_{\theta}}}\right]-i \mathrm{erfi}\left[\frac{1+i\beta k_{\theta} (\pi-\theta_{0} )}{\sqrt{2\beta k_{\theta}}}\right]\right)\right).
\end{gather}
\end{widetext}
This expression is real for positive and real values of $k_{\theta},\beta$ and $\theta_{0}$.
%\begin{equation}
%\frac{\sqrt{\pi}\left[erf\left(\theta_{ijk}^{0}\sqrt{\frac{\beta k_{ijk}^{\theta}}{2}}\right)-
%erf\left(\theta_{ijk}^{0}-\pi\right)\sqrt{\frac{\beta k_{ijk}^{\theta}}{2}}\right]}{2\sqrt{\frac{\beta k_{ijk}^{\theta}}
%{2}}}\end{equation}

When varying a dihedral angle, the potential term value depends on
the orientation of first bond (which determines the axis from which
the dihedral angle is measured). However, since the integration has
to be performed over all the range $\left[0,2\pi\right]$, integrating with respect to
the $\phi$ angle will yield a factor which is independent of the
location of the first bond. Thus, the integration does not depend on
the direction of the first bond and is straightforward.
For the commonly used dihedral angle potential we get
%\begin{equation}
%V_{d}\left(\phi_{ijkl}\right)=k_{\phi}\left(1+\cos\left(n\phi-\phi_{s}\right)\right).\nonumber \end{equation}
%The integration over this degree of freedom yields the following result:
\begin{eqnarray}
\label{eq:dihedral_angle}
Z_{d}&=&\int e^{-\beta k_{\phi}\left(1+\cos\left(n\phi-\phi_{s}\right)\right)}d\phi=2\pi e^{-\beta k_{\phi}}I_{0}\left(\beta k_{\phi}\right)\label{eq:dihedral},\end{eqnarray}
where $I_{0}$ is the modified Bessel function of the first kind. Note that this result does not depend on $n$, which is an integer.  In the context of restraints quadratic dihedral terms can be used and we give the exact expression for the corresponding partition function in the Appendix for completeness.   

It can now be seen that the partition function of transformed phenol in Eq. (1) can be decomposed (exactly) into the partition function associated with the $\mathbf{\Omega}$ degrees of freedom and the covalent bond, bond angle and dihedral angle partition functions.
\subsection*{Three or more Bonds Junctions}
Molecule shapes can include a monomer (covalent bond) that splits into more than one
monomer. Such examples are the trigonal planar, tetrahedral trigonal
pyramidal etc.  These cases will necessitate numerical integration which can be performed using the Spherical law
of cosines:
\begin{equation}
\cos\left(\theta_{12}\right)=\cos\left(\theta_{1}\right)\cos\left(\theta_{2}\right)+\sin\left(\theta_{1}\right)\sin\left(\theta_{2}\right)\cos\left(\Delta\phi\right)\label{eq:Spherical law of cosines},\end{equation}
where $\theta_{1},\theta_{2}$ denote the bond angles of two bonds
with respect to the principal bond and $\theta_{12},\Delta\phi$ denote
the bond angle and the difference in $\phi$ angle between these two
bonds respectively.
Usually in these cases there is one dihedral term, which we denote by $\phi_1$, that depends on the angle defined by
 one of the bonds, the principal bond and a previous bond.
Since the integration over the other degrees of freedom yields a factor
that is independent of the value of $\phi_1$, the integrations are decoupled.
Thus, the integration for the case of one monomer that splits into
two can be written as follows:
\begin{equation}
Z=\int e^{-\beta\left(V_{b}+V_{d}\right)}d\theta_{1} d\theta_{2}d\phi_{1}d\phi_{2}=Z_{d}Z_{3b},\end{equation}
where $Z_{d}$ is according to the definition in Eq. (\ref{eq:dihedral})  and
\begin{widetext}
\begin{equation}
Z_{3b}=
\int e^{-\frac{\beta}{2}\left[k_{1}^{\theta}\left(\theta_{1}-\theta_{1}^{0}\right)^{2}+k_{2}^{\theta}\left(\theta_{2}-\theta_{2}^{0}\right)^{2}+k_{12}^{\theta}\left(\theta_{12}-\theta_{12}^{0}\right)^{2}\right]}\sin(\theta_{1})\sin(\theta_{2})d\theta_{1}d\theta_{2}d\phi_{2}.
\end{equation}  
For the general case it can be written as follows:
\begin{equation}
Z_{nb}=\int\prod_{i}^{n}e^{-\frac{\beta}{2}k_{i}^{\theta}\left(\theta_{i}-\theta_{i}^{0}\right)^{2}}\prod_{i>j}e^{-\frac{\beta}{2}k_{ij}^{\theta}\left(\theta_{ij}-\theta_{ij}^{0}\right)^{2}}\prod_{i}^{n}\sin\theta_{i}d\theta_{i}\prod_{i\geq2}d\phi_{i},
\end{equation}
\end{widetext}
where $\theta_{ij}$ can be calculated from Eq. \eqref{eq:Spherical law of cosines} and $\phi_1$, which has to be substituted in $\Delta \phi$ in this equation, is an arbitrary constant.
In case there are energy terms that introduce complexity they can be relaxed in the transformation. For an explicit example with a bond junction see SM.

%To illustrate such a case we write the partition function of transformed Toluene (see Fig. 2), where as in the case of phenol, the Hamiltonian $H_r$ includes the improper dihedral, electrostatic and vdW potential terms which depend on the coordinates of the group of selected atoms
%
%\begin{gather}
%Z_{\mathrm{Toluene}'}=\int e^{-\beta H\left(\mathbf{\Omega}\right)}\prod_{i=1}^{3}e^{-\beta H_{r_{Hi}}(r_{Hi})}e^{-\beta H_{\theta_{Hi-C2}}(\theta_{Hi})}\times \nonumber \\
%e^{-\beta H_{\theta_{H1-H2}}(\theta_{H1-H2})-\beta H_{\theta_{H2-H3}}(\theta_{H2-H3})-\beta H_{\theta_{H3-H1}}\left(\theta_{H3-H1}\right)}\times \nonumber \\ 
%e^{-\beta H_{\phi_{H1}}\left(\phi_{H1}\right)}d\mathbf{\Omega} r_{Hi}^2 dr_{Hi}\sin\theta_{Hi}d\theta_{Hi}d\phi_{H1}. \nonumber
%\end{gather}
%
%\begin{figure}[h]
% \centering
%\includegraphics[width=8cm]{toluene}
% \caption{Transformed toluene}
% \label{fig:phenol}
%\end{figure}
%
%It can be readily seen that the partition function can be decomposed into the one associated with the $\mathbf{\Omega}$ degrees of freedom and the covalent bond, dihedral angle and three bonds junction partition functions.
%
The free energy factors can be substituted in:
\begin{equation}
\frac{\Delta F_{A'\rightarrow A''}}{k_{B}T}=\left(\sum_{i}\ln Z_{A_{c_{i}}}+\sum_{i}\ln Z_{A_{b_{i}}}+\sum_{i}\ln Z_{A_{d_{i}}}\right),
\end{equation}
to give the free energy difference, where $A''$ denotes a submolecule of the transformed molecule that is not necessarily realistic whose free energy will not be calculated.

In addition, the partition function of complex structures may be calculated numerically.
In many cases the complex structures need to be compared to identical ones, eliminating the need for these calculations.

Thus, we can write in terms of the partition functions:
\[
Z\rightarrow Z_{\mathrm{int}}\prod_{i=1}^{l}Z_{c_{i}}\prod_{i=1}^{m}Z_{d_{i}}\prod_{i=1}^{p}Z_{2b_{i}}\prod_{i=1}^{q}Z_{3b_{i}}\prod_{i=1}^{r}Z_{\mathrm{complex}_{i}},
\]
where $Z_{\mathrm{int}}$ represents the partition function
of the submolecule that is fully interacting and $Z_{c_{i}}$ and $Z_{d_{i}}$ represent the
$i$th covalent bond and dihedral angle partition function respectively.
$Z_{2b_{i}}$ and $Z_{3b_{i}}$ represent the $i$th two bond and
three or more bond junctions respectively and $Z_{\mathrm{complex}_{i}}$
represents the $i$th complex structure partition function. The arrow
represents the transformation $\lambda=1\rightarrow0$.

We can also choose the group of selected atoms as all the atoms of the molecule, and calculate the free energy in a similar manner.
The partition functions in this case can be written as follows:
\[
Z\rightarrow Z_{\mathrm{fp}} \prod_{i=1}^{l}Z_{c_{i}}\prod_{i=1}^{m}Z_{d_{i}}\prod_{i=1}^{p}Z_{2b_{i}}\prod_{i=1}^{q}Z_{3b_{i}}\prod_{i=1}^{r}Z_{\mathrm{complex}_{i}},
\] 
where $Z_{\mathrm{fp}}$ denotes the partition function of a free particle.

\begin{figure}[h]
 \centering
\includegraphics[width=8cm]{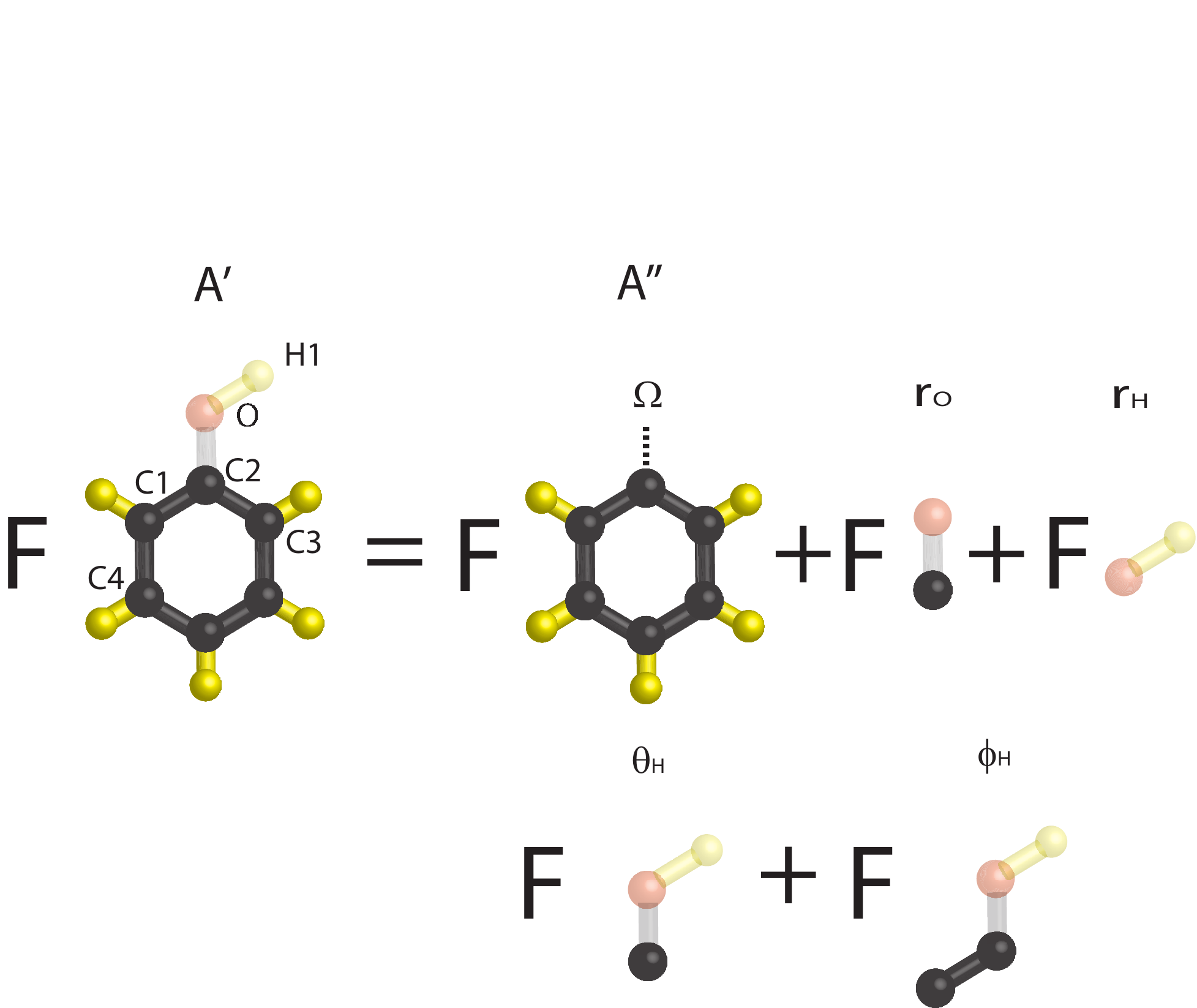}
 \caption{The decomposition of the free energies of the transformed molecule A'}
 \label{fig:phenol_free_energies}
\end{figure}

In Fig. \ref{fig:phenol_free_energies} we present the free energy decomposition of transformed phenol (A').  The degrees of freedom associated with each system are written on top. F denotes the free energy and A'' which is the reference (possibly imaginary) submolecule, is presented in the first term in the decomposition. It should be noted that the $\theta_{\mathrm{O}}$ and $\phi_{\mathrm{O}}$ degrees of freedom are associated with the submolecule A'' and the $r_{\mathrm{O}}$ degree of freedom is associated with the second system in the decomposition. The potential terms that depend on the coordinates of the $O$ and $H$ atoms in the five decomposed systems are: $1:V_{b}\left(\theta_{C1-C2-O}\right),V_{b}\left(\theta_{C3-C2-O}\right),V_{d}\left(\phi_{C4-C1-C2-O}\right)\,2:V_{c}\left(r_{O}\right),\,3:V_{c}\left(r_{H}\right),\,4:V_{b}\left(\theta_{C2-O-H1}\right)\,\, \mathrm{and}\,\, 5:V_{d}\left(\phi_{C1-C2-O-H1}\right)$. The reason that the degrees of freedom $\theta_{O}$ and $\phi_{O}$ are associated with the first system in the decomposition is that the corresponding potential terms usually do not depend on the atom (e.g O).  
The free energies of the second and third systems are calculated according to Eq. (\ref{eq:covalent_bond}) and the free energies of the fourth and fifth systems are calculated according to Eqs. (\ref{eq:bond_angle}) and (\ref{eq:dihedral_angle}) respectively.
%Thus, comparison between any group of reactants and product can be performed by transforming each molecule in a separate simulation, followed by calculation of the free energies associated with the different submolecules.
%Since the relaxed vdW and electrostatic terms involve diverging terms at $r \rightarrow 0$, even at $\lambda \rightarrow 0$ these terms will still be dominant. Thus, in order for the calculations to be legitimate, soft core potentials have to be used (see for example Ref. \cite{farhi2013simple}).
%In the case of rugged energy landscape, sampling techniques such as H-REMD have to be used in another $\lambda$ dimension. In this free energy calculation method H-REMD can be used in the same $\lambda$ dimension \cite{farhi2013simple}.
\section*{Demonstrations}
The covalent bond free energy calculation has been demonstrated and compared with numerical integration for two molecules of two atoms in a spherical potential (see Supplementary Material). In addition, for methanethiol molecule the free energies were calculated according to  Eqs. (2)-(7) and were in agreement with the results obtained by performing transformations in MD simulations. The computations using these equations were faster by factors varying between $5\cdot10^{3}$ and $10^{12}$ (for the same computational resources) \cite{ToBePublished} . 
%We write the Hamiltonian in its final form:
% \begin{equation}
%H'_{A}\left(\lambda\right)=\lambda
%H'_{_{{A_{dc}/B_{dc}}}}+f(\lambda) H_{{}{A}}
%\end{equation}
%where $H_{_{{A_{dc}/B_{dc}}}}$ denote the coupling terms of the different sub molecule.
\section{potential application}
Here we suggest the potential application of calculating free energies of chemical reactions using classical molecular simulations followed by analytic or numerical calculations. We will calculate the free energies of only small parts of the molecules, possibly in solvent, to obtain the free energy of the chemical reaction which can involve large molecules. 

To obtain the equilibrium constant of a chemical reaction the standard Gibbs free energies (which are similar to Helmholtz free energies in many cases \cite{abraham2015gromacs}) are usually calculated. The standard state is the hypothetical state with the standard state concentration but exhibiting infinite-dilution behavior (the interactions between e.g the solute molecules are negligible). Hence, when we are interested in the standard free energy of one substance \emph{a single copy of the molecule of interest can be simulated} either in vacuum or solvent environments.

\begin{figure}[h]
 \centering
\includegraphics[width=8cm]{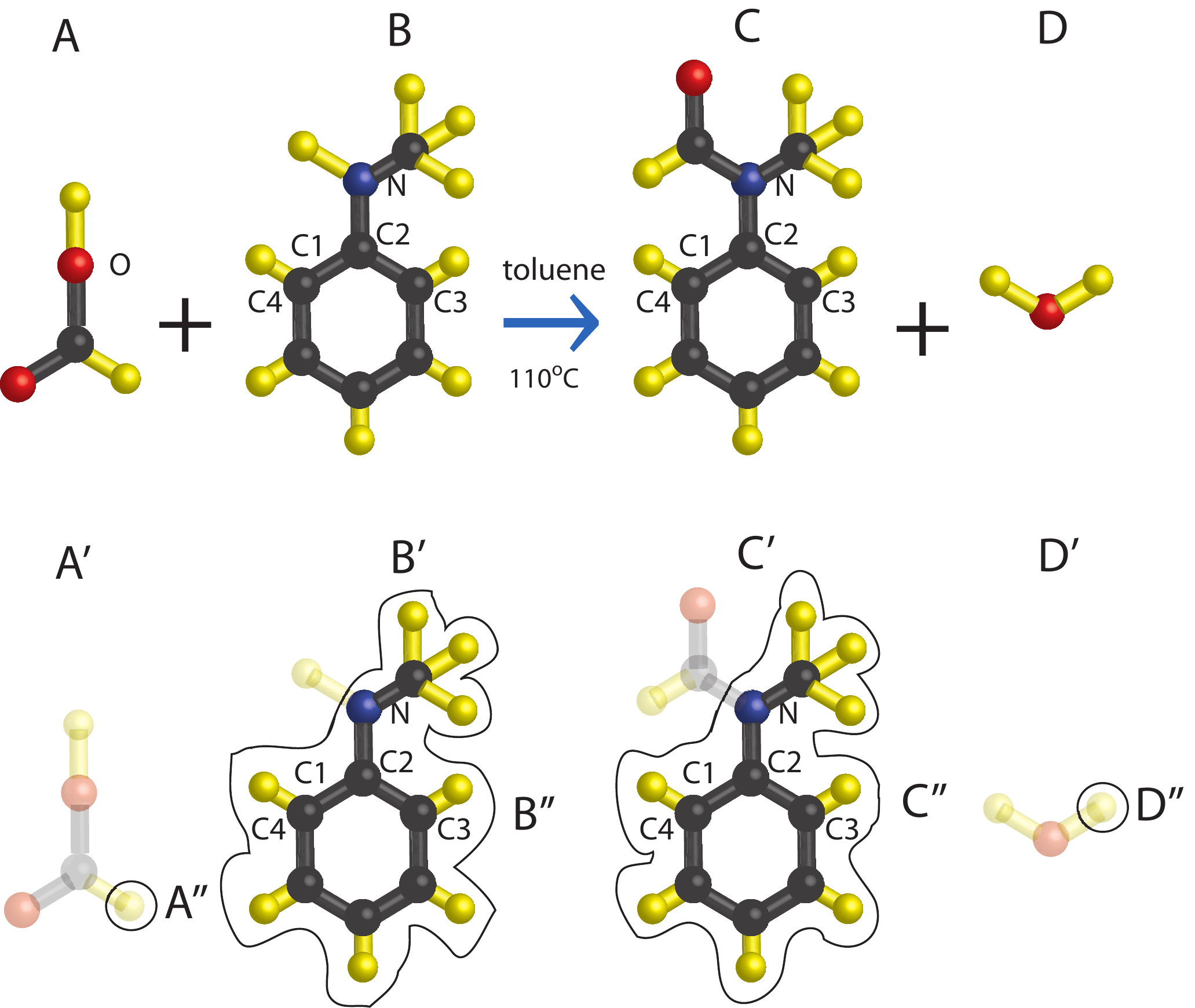}
 \caption{A scheme of the free energy calculation of the chemical reaction: Formic acid + N-methylaniline $\rightarrow$ N-Methylformanilide+water}
 \label{fig:chemical_reaction}
\end{figure}

The idea is to transform the reactants and the products, between which
the free energy difference is calculated, into molecules that
have the same partition functions up to factors that can be calculated.
First, we match reactant molecules with product molecules that have submolecules in common if possible so that the free energy of these submolecules will cancel out. 
Then, we transform the molecules by relaxing potential terms of the atoms that are different between the molecules and calculate the free energy differences associated with each transformation. Finally the free energy factors associated with the different atoms are calculated and we can deduce the free energy of the chemical reaction.
For example, given the chemical reaction
\begin{equation}
 A+B\rightarrow C+D,
\end{equation}  
we can match, if possible, molecules $A,B$ to molecules $C,D$. Then we transform each of the molecules to  $A',B',C'$ and $D'$ and calculate the free energy differences $\Delta F_{A\rightarrow A'},\Delta F_{B\rightarrow B'},\Delta F_{C\rightarrow C'}$ and $\Delta F_{D\rightarrow D'}$ respectively. Finally we calculate the free energy factors $\Delta F_{A'\rightarrow A''},\Delta F_{B'\rightarrow B''},\Delta F_{C'\rightarrow C''}$ and $\Delta F_{D'\rightarrow D''}$ as previously explained. Thus we get
\begin{widetext} 
\begin{equation}
\Delta F_{A+B \rightarrow C+D}=\Delta F_{A\rightarrow A'}+\Delta F_{B\rightarrow B'}-\Delta F_{C\rightarrow C'}-\Delta F_{D\rightarrow D'}+\Delta F_{A'\rightarrow A''} +\Delta F_{B'\rightarrow B''} -\Delta F_{C'\rightarrow C''} -\Delta F_{D'\rightarrow D''}.
\end{equation}
\end{widetext}
An example for such a calculation is given in Fig. \ref{fig:chemical_reaction}. The molecules that participate in the chemical reaction and the transformed molecules are presented on top and bottom respectively. In this case molecule B is matched with C and $F_{B''}=F_{C''}$. $F_{A''}$ and $F_{D''}$, which are the free energy of a free particle, are equal. 

We presented an accurate and complete method for calculating molecular free energies in classical potentials. We suggested the potential application of free energy calculation of chemical reactions in classical potentials. This application can find use in organic chemistry, biochemistry and drug discovery.
\subsection*{Acknowledgements}
D.J. Bergman, A. Nitzan, D. Harries, G. Falkovich and B. Singh are acknowledged for the useful comments.
%\section*{Appendix}
{

\subsection*{Appendix}
The exact expression for the partition function associated with a quadratic dihedral term is the following 
$$Z_{d\,\mathrm{quadratic}}=\int e^{-\beta k_{d}\left(\phi-\phi_{0}\right)^{2}}d\phi=\nonumber$$ 
$$\frac{\sqrt{\pi}\left[\text{erf}\left(\phi_{0}\sqrt{\beta k_{d}}\right)-\text{erf}\left(\left(\phi_{0}-2\pi\right)\sqrt{\beta k_{d}}\right)\right]}{2\sqrt{\beta k_{d}}}.$$
  
\bibliographystyle{apsrev4-1}
\bibliography{bib12}

%\end{spacing}
\end{document}